\documentclass[twocolumn,prl,preprintnumbers,showpacs,aps,amssymb]{revtex4}

\usepackage{graphicx}
\usepackage{bm}
\usepackage{amsmath}

\def\calL{{\cal L}}

\def\calU{{\cal U}}

\def\BZ{{\cal BZ}}

\def\hbar{{\bar h}}

\def\qbar{{\bar q}}

\def\calUslash{\calU\hspace{-1.9mm}/}
\def\calUnot{\calU\hspace{-2.4mm}/}

\def\nn{\nonumber}

\begin{document}
\title{Unparticles and holography}
\author{Jong-Phil Lee}
\email{jplee@kias.re.kr}
\affiliation{Korea Institute for Advanced Study, Seoul 130-722, Korea}
\preprint{KIAS-P07070}

\begin{abstract}
We construct the holographic dual theory of unparticles.
The Randall-Sundrum type hard wall model is shown to produce deconstructing 
particles, whose spectrum has a finite mass gap proportional to the inverse of
the fifth direction segment.
The introduction of new scale corresponds to setting a brane in ${\rm AdS}_5$.
The broken conformal symmetry due to this brane is restored when it is moved 
to infinity.
Unparticles then emerge as an infinite tower of massless particles.
\end{abstract}

\maketitle
Scale invariance plays a useful role in a variety of fields in physics,
from phase transitions to string theory.
The bizarre feature of the scale invariant sector was studied many years ago
by Banks and Zaks($\BZ$) in the context of gauge theory \cite{BZ}.
They examined the infrared(IR)-stable fixed point of Yang-Mills theories with
massless fermions,
and found that for a proper number of fermions in a certain representation
the theory is chirally invariant and has no mass gap.
\par
But in low energy particle physics, we have been confronted by the
zoo of particles with a very wide mass spectrum, which explicitly
breaks the scale invariance. Thus the scale invariant sector
cannot be thought as ordinary particles. Recently, Georgi proposed
a possibility that such a scale invariant stuff might exist and
remain hidden, with the name of "unparticles" \cite{Georgi}. At
some high energy, there are $\BZ$ fields with a nontrivial IR
fixed point. There existence and interactions with the standard
model(SM) particles are nicely described by an effective theory. The
effective theory involves the unparticle operators on which the
$\BZ$ operators are matched. For an unparticle operator of scaling
dimension $d_\calU$, the unparticle appears as a non-integral
number $d_\calU$ of invisible massless particles. After the
Georgi's suggestion, there have been a lot of phenomenological
studies on unparticles \cite{Cheung}.
\par
Another triumph involving the scale invariance during the decade
is the AdS/CFT correspondence \cite{Maldacena,Gubser,Witten:1998qj}.
According to this, a strongly
coupled conformal field theory  on 4D is equivalent to a gravity
theory on AdS in 5D. Recent works include the construction of 5D
theory equivalent to the 4D QCD, which have been very successful.
The essence of the equivalence is that for every CFT operator
there exists a corresponding field in AdS. We assume that this
correspondence still holds for the scale invariant $\BZ$ sector
\cite{Stephanov}. Since the most fundamental feature of the
unparticle is the scale invariance, it is quite natural to try to
apply the AdS/CFT correspondence and to figure out the holographic
dual description of the unparticle in AdS. Once we accept the
correspondence, then for an unparticle operator there must be a
corresponding bulk field in 5D AdS. In this respect, the bottom-up
approach for the AdS/QCD is a good benchmark.
\par
Another reason for considering the AdS/QCD is that the unparticle
operators are matched from the $\BZ$ operators which originate
from the massless fermions of the non-Abelian gauge theory. The
theory is basically QCD, except the asymptotic freedom. Thus we can
inherit many developments of the holographic QCD. What makes the
theory different from QCD is the scale invariance.
\par
As an example, consider a QCD operator $\qbar q$. According to the
AdS/CFT correspondence, there is a bulk field $\phi$ in the
holographic dual theory in ${\rm AdS}_5$. Its 5-dimensional mass
is given by, in general $m_5=(\Delta-p)(\Delta+p-4)$, where
$\Delta$ is the canonical dimension of the 4D $p$-form operator.
And it is quite well known that the bulk field $\phi$ scales
$\phi(x,z)\sim c_1 z^{4-\Delta}+c_2z^\Delta$ for $z\to 0$ where
$z$ is the fifth dimension \cite{KW}. For $\qbar q$, $m_5=-3$ and
$\phi\sim c_1z+c_2z^3$.
Similarly for an unparticle (scalar) operator of a non-integer dimension
$d_\calU$ the 5D mass is $m_5=d_\calU(d_\calU-4)$, and the 5D
field scales with non-integral power of $z$.
\par
In addition, the chiral symmetry of the $\BZ$ sector in 4D is much like that
of the massless QCD.
In the holographic dual theory, there is a corresponding local gauge symmetry
from which the 5D vector gauge bosons are naturally defined.
The 4D object, dual to the 5D vector gauge boson, can be identified as the $\BZ$
vector operator.
\par
The holographic dual picture of AdS/QCD can be well explained by a
simple Randall-Sundrum(RS) \cite{RS} type setup
\cite{ArkaniHamed:2000ds,HW,Da Rold:2005zs}. Especially in the
RS1-type background, there are two branes at $z=\epsilon$ and
$z=z_c$. The quantity $1/z$ plays a role of the renormalization
scale in 4D. The brane located at $z=\epsilon(\to 0)$ (UV brane)
puts a UV cutoff ($\sim 1/\epsilon$) in the 4D theory, while the
one at $z=z_c$ (IR brane) determines the typical mass scale of
QCD, $\Lambda_{\rm QCD}\sim 1/z_c$.
\par
A very similar correspondence can also be applied to the unparticles.
In the framework of \cite{Georgi}, there are two relevant scales in the
unparticle physics.
One is a very high energy scale $M_\calU$.
The SM fields and the $\BZ$ fields interact via the exchange of heavy
particles with the large mass scale $M_\calU$.
Thus the scale $M_\calU$ is the UV cutoff where a new physics appears.
Below the scale $M_\calU$, the interactions between the SM and $\BZ$ fields
are described by the nonrenormalizable couplings suppressed by powers of
$M_\calU$.
Then through the dimensional transmutation the renormalizable couplings of
the $\BZ$ fields induce another scale $\Lambda_\calU$ where the
scale-invariant $\BZ$ sector appears.
Below the scale $\Lambda_\calU$ we construct an effective theory in the
context of unparticle operators onto which the $\BZ$ operators match
at $\Lambda_\calU$.
Since the theory enters the conformal regime at $\Lambda_\calU$,
we build a brane in 5D dual theory located at $z=z_0\sim 1/\Lambda_\calU$.
In the region $z>z_0$, one can expect the AdS/CFT correspondence works well
because the 4D theory is really conformal with the unparticle operators.
We will call this a "$\calU$ brane".
\par
Recently an interesting picture for unparticles is suggested \cite{Stephanov}.
Here the unparticles are "deconstructed" to be an infinite tower of particles
of different masses.
There is a mass gap of order $\Delta_m$ which will be sent to zero.
The true unparticles appear when $\Delta_m$ vanishes.
Because the nonvanishing parameter $\Delta_m$ introduces a scale in the theory,
one may think of $\Delta_m$ as a scale-invariance breaking,
or non-conformal scale $\Lambda_\calUslash$.
In a 5D description, this non-conformal scale corresponds to an IR brane
located at $z=z_m\sim 1/\Delta_m\sim 1/\Lambda_\calUslash$.
We call this a "$\calUnot$ brane".
From the study of AdS/QCD, it is well known that this kind of hard-wall model
produces the mass spectrum $m_n^2\sim n^2/z_m^2$.
Now we expect that a similar mass spectrum occurs in the
$z_0\sim 1/\Lambda_\calU<z<z_m\sim 1/\Lambda_\calUslash$ model.
The existence of the $\calUnot$ brane at $z=z_m$ represents a departure from 
AdS.
The corresponding effect on 4D theory is the breaking of a scale invariance
for $z>z_0$, which means a nonzero mass gap
($\sim 1/z_m\sim \Lambda_\calUslash$) in the spectrum.
But the scale invariance is restored when $z_m\to\infty$ with the vanishing
mass gap $\Lambda_\calUslash\to 0$,
and the infinite tower of massless particles are identified as unparticles.
\par
We start with the simple 5D AdS metric of RS type,
\begin{equation}
ds^2=\frac{1}{z^2}\left(dx^\mu dx_\mu-dz^2\right)~,
\end{equation}
with the $\calU$ and $\calUnot$ branes at $z=z_0$ and $z=z_m$ respectively.
In the 5D bulk there are left(right)-handed gauge bosons
$A_{L(R)}=A_{L(R)}^a t^a$ of the gauge symmetry 
${\rm SU}(3)_L\otimes{\rm SU}(3)_R$, 
and a scalar field $\Phi=S\exp(i2\pi^a t^a)$,
where $t^a$ are the ${\rm SU}(N)$ generators.
Here $S$ is a real scalar while $\pi^a$ are real pseudoscalars.
The 5D action of these fields is given by
$S_5=\int d^4x\int dz\calL_5$ with
\begin{equation}
\calL_5=\sqrt{g}{\rm Tr}\left\{
-\frac{1}{4g_5^2}(F_L^2+F_R^2)+|D\Phi|^2-M_\Phi^2|\Phi|^2\right\}~,
\end{equation}
where $g={\rm det}(g_{MN})$, $D_M=\partial_M\Phi-iA_{LM}\Phi+i\Phi
A_{RM}$ ($M=\mu,z$) and $F_{MN}=\partial_M A_N-\partial_N A_M -i[A_M,A_N]$.
Here $g_5$ is the 5D gauge coupling which can be matched onto the
4D theory to be $g_5^2=12\pi^2/N_c$ where $N_c$ is the number of
color \cite{HW}.
\par
Note that the 5D mass of the vector gauge boson in the holographic
QCD vanishes for $\Delta=3$ and $p=1$. It means that the local
gauge symmetry in 5D is preserved. But for a vector boson
corresponding to the vector unparticle operator with dimension
$d_\calU$, its 5D mass will not vanish in general:
$M_V^2=(d_\calU-1)(d_\calU-3)$.
We assume that $1<d_\calU<2$ \cite{Georgi}.
Thus in the 5D holographic model
of unparticles the gauge symmetry is broken, though we do not know
the details about the breaking mechanism. We now construct the 5D
Lagrangian with the gauge boson mass term:
\begin{equation}
\calL_5^\calU=\sqrt{g}{\rm Tr}\left\{-\frac{1}{4g_5^2}
(2F_V^2-4M_V^2V^2)+|D\Phi|^2-M_\Phi^2|\Phi|^2\right\}~,
\end{equation}
where $V_M=(A_L+A_R)/2$.
We choose $V_z(x,z)=0$ gauge.
Here the possible axial vector part is dropped.
Fluctuations around the vacuum expectation value of $S$ give
the real scalar degree of freedom $\sigma$: $S=v(z)/2+\sigma$. The
equation of motion for $\sigma$ is
\begin{equation}
\left(\partial_z^2-\frac{3}{z}\partial_z
+m_S^2-\frac{M_\Phi^2}{z^2}\right)\sigma=0~.
\end{equation}
The general solution is
\begin{equation}
\sigma(z)=c_1z^2J_{d_\calU-2}(m_Sz)+c_2z^2Y_{d_\calU-2}(m_Sz)~.
\end{equation}
At $z=z_0$, we impose $\sigma(z_0)=0$ as a boundary condition 
\cite{HW,Da Rold:2005zs}.
The wavefunction becomes
\begin{eqnarray}
\sigma(z)&=&c_1z^2\left[J_{d_\calU-2}(m_Sz)
 -\frac{J_{d_\calU-2}(m_Sz_0)}{Y_{d_\calU-2}(m_Sz_0)}Y_{d_\calU-2}(m_Sz)
\right]\nn\\
&\to&
c_1z^2J_{d_\calU-2}(m_Sz)~~~{\rm as}~z_0\to 0~.
\end{eqnarray}
From the boundary condition at $z=z_m$, the mass spectrum of $m_S$ is
obtained. 
We require $\sigma'(z=z_m)=0$.
For large $m_Sz_m\gg 1$, the boundary condition leads to
$J_{d_\calU-1}(m_Sz_m)\simeq 0$, i.e.,
\begin{equation}
m_{S,n}\simeq\frac{\pi}{z_m}\left(\frac{2d-3}{4}+n\right)~.
\label{mS}
\end{equation}
\par
Similarly, the equation of motion for the vector boson is
\begin{equation}
\left(\partial_z^2-\frac{1}{z}\partial_z
+m_V^2-\frac{M_V^2}{z^2}\right)V_M=0~.
\end{equation}
The solution is
\begin{equation}
V_M(z)=c_3zJ_{d_\calU-2}(m_Vz)+c_4zY_{d_\calU-2}(m_Vz)~.
\end{equation}
Imposing the same boundary condition, $V_M(z_0)=V_M'(z_m)=0$,
gives the same mass spectrum of $m_V$ as $m_S$, though the
equation of motion is slightly different:
\begin{equation}
m_{V,n}\simeq\frac{\pi}{z_m}\left(\frac{2d-3}{4}+n\right)~.
\label{mV}
\end{equation}
The mass spectra (\ref{mS}) and (\ref{mV}) are what we have expected before.
The behavior of $m_n\sim n/z_m$ is the same as that of \cite{Stephanov}.
For a finite value of $z_m$, the spectra of $m_{S,n}$ and $m_{V,n}$ are 
discrete and the mass gap is finite.
The scale $\sim 1/z_m$ explicitly breaks the conformal symmetry.
Note that if $d_\calU=3$, $m_{V,n}$ reproduces the well known results for
the vector meson spectrum \cite{Da Rold:2005zs}.
The unparticles emerge in the limiting situation where $z_m\to\infty$;
the scale invariance is restored and there are only infinite tower of
massless particles.
In this sense, the RS2-type setup is a very good holographic dual theory of 
unparticles in ${\rm AdS}_5$.
\par
In conclusion, we have constructed the holographic dual theory of unparticles.
The emergence of the scale invariance of the $\BZ$ fields at $\Lambda_\calU$
corresponds to the $\calU$ brane in the RS background.
For scalar and vector unparticle operators, there are bulk fields $\Phi$ and
$V_M$ much like those in the holographic QCD.
The introduction of a conformal symmetry breaking scale $\Lambda_{\calUslash}$
is equivalent to putting a $\calUnot$ brane in 5D at 
$z_m=1/\Lambda_{\calUslash}$.
This is also a scale of mass gap of the 4D spectra for $\Phi$ and $V_M$.
The scale invariance is recovered when $z_m\to\infty$ and the unparticles are
identified as the infinite tower of massless particles.
The limiting case of putting $\calUnot$ brane at infinity is equivalent to the
RS2 scenario, which is a good holographic description of the unparticles.
\par
\begin{acknowledgments}
The author thanks Youngman Kim and Taeil Hur for helpful discussions.
\end{acknowledgments}

\end{document}